\documentclass[journal=jpccck,manuscript=article]{achemso}

\usepackage{url}
\usepackage{graphicx}
\usepackage{dcolumn}
\usepackage{bm}
\usepackage{amssymb,amsmath}

\title{Long-timescale simulations of H$_2$O admolecule diffusion on Ice Ih(0001) surfaces}

\author{Andreas Pedersen} 
\affiliation{Faculty of Physical Sciences and Science Institute, University of Iceland, 107 Reykjav\'{\i}k, Iceland}
\affiliation{Integrated Systems Laboratory, ETH Zurich, 8092 Zurich, Switzerland}

\author{Leendertjan Karssemeijer} 
\affiliation{Radboud University Nijmegen, Institute for Molecules and Materials, Heyendaalseweg 135, 6525 AJ Nijmegen, The Netherlands}

\author{Herma M. Cuppen}  
\affiliation{Radboud University Nijmegen, Institute for Molecules and Materials, Heyendaalseweg 135, 6525 AJ Nijmegen, The Netherlands}

\author{Hannes J\'onsson}
\affiliation{Faculty of Physical Sciences and Science Institute, University of Iceland, 107 Reykjav\'{\i}k, Iceland}
\affiliation{Department of Applied Physics, Aalto University, Espoo, FI-00076, Finland}
\email{hj@hi.is}

\begin{document}

\begin{abstract}
Long-timescale simulations of the diffusion of a H$_2$O admolecule on the (0001) basal plane of ice Ih
were carried out over a temperature range of 100 to 200 K using the adaptive kinetic Monte Carlo method 
and TIP4P/2005f interaction potential function. 
The arrangement of dangling H atoms was varied from the proton-disordered surface to the perfectly ordered Fletcher surface. 
A large variety of sites was found leading to a broad distribution in adsorption  
energy at both types of surfaces. Up to 4\% of the sites on the proton-disordered surface
have an adsorption energy exceeding the cohesive energy of ice Ih.  
The mean squared displacement of a simulated trajectory at 175 K for the proton-disordered surface gave a diffusion constant of 
6$\cdot$10$^{-10}$ cm$^2$/s, consistent with an upper bound previously reported from experimental measurements.
During the simulation, dangling H atoms were found to rearrange so as to reduce clustering, 
thereby approaching a linear Fletcher type arrangement.
Diffusion on the perfectly ordered Fletcher surface was estimated to be significantly faster, especially in the direction along the rows
of dangling hydrogen atoms.
From simulations over the range in temperature, an effective activation energy of diffusion was estimated to be
0.16~eV and 0.22~eV for diffusion parallel and perpendicular to the rows, respectively.
Even a slight disruption of the rows of the Fletcher surface made the diffusion isotropic. 
\end{abstract}

\maketitle

\section{Introduction}
\label{sec:Introduction}

In the most common form of ice on Earth, the hexagonal Ih structure, the oxygen atoms of the water molecules reside in a hexagonal lattice. 
Each O atom forms four bonds in a tetrahedral arrangement with neighboring H atoms, two covalent bonds and
two hydrogen bonds. One and only one H atom sits in between each pair of neighboring O atoms.
As long as these {\it ice-rules}~\cite{Bernal33} are obeyed, 
the arrangement of the H atoms is rather arbitrary while the O atoms sit on a regular lattice.

Thermal energy He-atom scattering experiments~\cite{Glebov00}  as well as low-energy electron diffraction (LEED)~\cite{Materer97} indicate a full-bilayer termination of the (0001) surface at low temperature where the surface is not premelted, 
although the outermost molecules were not detectable by LEED because of large vibrational amplitudes~\cite{Materer97}. 
When the proton-disordered ice Ih crystal is cut in between bilayers to create two surfaces, 
some of the hydrogen atoms of the surface molecules will not form hydrogen bonds.
These dangling H atoms (DH) will form a disordered pattern on the surface. 
There is, however, a repulsive dipole-dipole interaction between the DHs and a lower energy
ordering involves formation of rows where each DH has only two neighboring DH. This reduces the repulsion between the dipoles.  
Regular rows of DH on the surface of the otherwise proton-disordered ice was predicted by Fletcher to have lower energy~\cite{Fletcher92}.  
Examples of the two types of surface structures are shown in Fig.~1.  
The configurational entropy of the Fletcher surface is low and at finite temperature
one can expect the rows to become disordered to some extent.
Buch {\it et al.} proposed that the DH form a mosaic with an intermediate degree of 
linear order~\cite{Buch08} to reach a compromise between increased entropy and reduced surface energy.  
The He-atom scattering measurements~\cite{Glebov00} indeed revealed small features in the reflected intensity which 
have been interpreted in terms of regularly spaced Fletcher type rows~\cite{Buch08}.
Recently, DFT calculations have lent some support for the stability of the DH ordering proposed by Fletcher~\cite{Pan10}.

The adsorption and diffusion of water admolecules on the ice surface is of central importance in the modeling
of ice crystal growth, which in turn is important for predicting the shape and properties of ice particles.
The diffusivity of water admolecules on the ice Ih(0001) surface at low temperature was studied experimentally by 
Brown and George~\cite{Brown96} who placed an upper bound of 5$\cdot$10$^{-9}$ cm$^2$/s on the diffusion constant at 140 K.
Batista and J\'onsson (BJ) carried out simulations of admolecule binding and diffusion on a 
proton-disordered surface using the TIP4P
interaction potential and the nudged elastic band method for finding diffusion paths~\cite{Batista01,Batista99} 
between sites identified by local energy minimization from random starting configurations. 
They pointed out that the DH disorder at the surface leads to a wide variety of adsorption sites, 
giving rise to a broad distribution of both the admolecule binding energy and the activation energy for diffusion hops between sites. 
The sites can be broadly classified in categories depending on how many DH 
are on the three underlying H$_2$O surface molecules. The 
strongest binding is obtained for sites with one or two DH, while sites with no DH bind only weakly and sites with three DHs 
do not provide stable binding.
The disordered arrangement of the two types of sites that bind most strongly leads to irregular diffusion paths on the surface.
Approximate kinetic Monte Carlo (KMC) simulations were carried out by BJ where the binding energy of sites was assigned
the average value for that type of site and the activation energy was drawn from a calculated distribution of 
energy barriers while the prefactor was assumed to be 10$^{12}$ s$^{-1}$. 
From the temperature dependence of the simulated diffusivity, 
an effective activation energy for diffusion was estimated to be 0.18 eV and the diffusion constant 
1$\cdot$10$^{-9}$ cm$^{2}$/s at 140 K, on the order of the upper bound placed by Brown and George from
experimental measurements.

Nie, Bartelt and Th\"urmeron measured the ripening of islands on 4-5 nm thick ice films on a Pt(111) surface and
interpreted the results in terms of water admolecule diffusion from small to large islands. 
From measurements carried out in a temperature range from 115 to 135 K, a combined activation energy for formation and 
diffusion of admolecules was determined to be 0.3 to 0.5 eV~\cite{Nie09}.  
If the calculated estimate of the average diffusion activation energy of BJ is used, 
the formation energy can be estimated to be on the order of 0.1 to 0.3 eV~\cite{Nie09}.
Due to the complexity of the system, it is not clear how to interpret this quantity. 

In the present paper, the results of more accurate simulations than the ones carried out by BJ are presented.
They are based on the 
adaptive kinetic Monte Carlo method~\cite{Henkelman01} (AKMC) where the 
mechanism and rate of diffusion events, as well as annealing events, is found from randomly initiated saddle point searches.
The binding energy of the various sites found from the diffusion simulation as well as activation energy 
and pre-exponential factor obtained for each diffusion event are included without the averaging used by BJ. 
The AKMC method has previously been used, for example, 
in simulation studies of the structure of grain boundaries and their effect on H atom diffusion~\cite{Pedersen09a,Pedersen09b}.
A more sophisticated description of the molecular interactions was also applied, 
the flexible TIP4P/2005f potential function~\cite{Gonzalez11}
instead of the original TIP4P where the geometry of the water molecules is frozen.
Furthermore, the diffusion was simulated for five different models of the ordering of the DHs on the surface.

The AKMC method is a powerful method for simulating long-timescale dynamics of complex systems such as ice Ih where
the proton disorder makes the simulation particularly challenging.  
A study of Ih(0001) surface annealing based on AKMC simulations was recently presented~\cite{Pedersen14}.
The applicability of the AKMC method 
for studies of molecular diffusion on ice surfaces was also illustrated recently by simulations of CO diffusion on the
proton-disordered ice Ih(0001) surface~\cite{Karssemeijer12}.

In the following section, Section II, the simulation methodology and surface models are presented. 
The simulation results are presented in section III. 
The article concludes with a discussion and summary in Section IV.
 

\section{II. Methodology}
\label{sec:Methodology}

\subsection{Surface models}

Five models of the Ih(0001) surface with different surface DH patterns were constructed. 
All samples contain 360 water molecules, arranged in 6 bilayers. The $z$-axis is chosen to lie along the $c$-direction and 
periodic boundary conditions are applied in the plane of the surface, the $x-y$ plane, to mimic an infinite slab. 
The construction of each sample started from a three-dimensional ice crystal with proton disorder 
generated using the method of Buch {\it et al.}~\cite{Buch98}. 
The net dipole of the samples is zero.
The atomic coordinates, as well as $a$ and $c$ lattice constants, were optimized to minimize the energy. 
To create the surface, two bilayers 
were frozen in the bulk configuration, a vacuum layer inserted below, and the four movable bilayers relaxed. 
The relaxed surface generated this way will be referred to as the disordered surface. To create a Fletcher surface, 
the same method was applied for generating a proton-disordered structure except that the
hydrogen bonds between two of the bilayers where made to form rows. 
These two bilayers were then separated and a vacuum layer inserted. 
Similarly, three samples of slightly disordered Fletcher surfaces with broken rows were created, as illustrated in Fig.~1.
From previous simulations of CO molecule diffusion on ice surfaces \cite{Karssemeijer12,Karssemeijer14}, where the effect of
system size was tested, we expect the surface models used here to be large enough to estimate the diffusivity.

\begin{figure}[t]
\centering
\includegraphics[width=.65\textwidth]{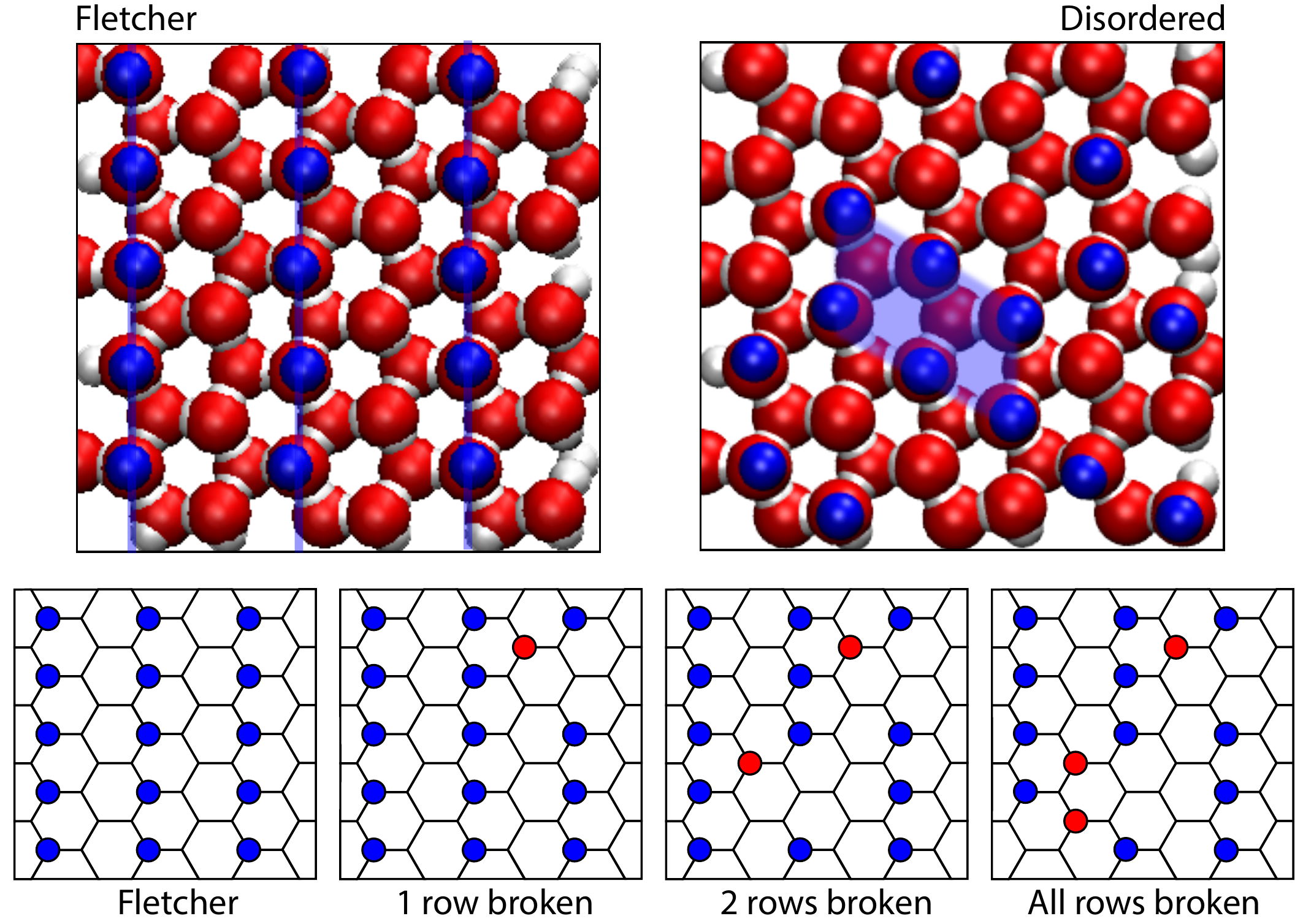}
\caption{Patterns of dangling H atoms (colored blue) on ice Ih(0001) surfaces used in the simulations.
Upper panel: Fletcher surface and surface with disordered arrangement of dangling H atoms showing clustering 
in the central region. 
Red spheres indicate O atoms, white spheres H atoms that participate in hydrogen bonding.
Lower panel: Varying degree of disorder added to the Fletcher surface.  Blue disks mark dangling H atoms arranged as in the Fletcher surface, red disks denote dangling H atoms disrupting the linear ordering.  
The degree of deviation from the Fletcher surface is characterized in terms of the number of broken rows.  
}
\label{fig:InitialSurfaces}
\end{figure}

The inter- and intramolecular interactions were modeled with the TIP4P/2005f potential~\cite{Gonzalez11}, which is a flexible version of the TIP4P/2005 potential~\cite{Abascal05}.  
In a previous study of surface annealing events this potential function was found to give results in good correspondence with DFT calculations \cite{Pedersen14}.
Periodic boundary conditions were applied but the molecular interactions 
were smoothly brought to zero when the centers of mass of the molecules were separated by 9 to 10~\AA. 
The relaxed, proton-disordered crystal has an $a$ to $c$ ratio of 1.738, which is 6.4\% larger than the ratio for a perfect HCP lattice 
and 6.8\% larger than the experimentally observed ratio for ice Ih~\cite{Rottger12}. 

The energy of the slab with the proton-disordered surfaces is highest, 46 meV larger 
than the DH ordered Fletcher surface \cite{Pedersen14}. 
The slab where one DH row on the surface is broken, shown in Fig.~1, turns out to have the lowest energy, slightly
lower than the perfect Fletcher surface.  
This illustrates the influence of the long range electrostatic field from the proton-disordered crystal slab which makes the
sites on the Fletcher surface inequivalent even though the surface DH rows are ordered.


\subsection{Diffusion simulation}

The AKMC method~\cite{Henkelman01} was used to simulate the diffusion of a single H$_2$O admolecule  
on the ice surfaces described above.  
The simulation starts from an energy minimized configuration where the admolecule is initially placed at a random 
point above the surface.
A path is generated consisting of a sequence of local minima on the energy surface corresponding to
adsorption sites on the substrate and first order saddle points representing transition states for diffusion hops between sites.
This path represents a possible time evolution of the system over a time interval that is much longer than what 
could be simulated with direct classical dynamics including vibrational motion of the atoms.
The EON software~\cite{Pedersen10,Chill14} was used to conduct the simulations.  
For each local minimum, several searches for low lying first order saddle points were carried out 
using the minimum-mode following (MMF) method~\cite{Henkelman99,Olsen04,Pedersen11}. 
There, the eigenmode of the Hessian matrix corresponding 
to the lowest eigenvalue -- the minimum mode -- is used to transform the force acting on the atoms in such a way that the vicinity of a first order saddle point becomes analogous to that of an energy minimum. 
By inverting the force component parallel to the minimum mode, a climb up the 
potential energy surface and convergence onto a first order saddle point can be conducted by applying an ordinary minimization 
algorithm where the gradient of the objective function is zeroed.  
The minimum vibrational mode was estimated here using the Lanczos method~\cite{Olsen04,Tennyson86,Malek00}. 
After locating a saddle point, the two adjacent minima were found by displacing the 
system slightly along and opposite to the direction of the minimum-mode eigenvector at the saddle point, 
followed by energy minimization. 
Searches for saddle points 
and minima were considered converged when the maximum force acting on any of the atoms dropped below 1~meV/\AA. 
The thermal transition rate due to trajectories passing through the vicinity of each of the saddle points was estimated using 
harmonic transition state theory (HTST)
\begin{eqnarray}
\label{equ:HTSTRate}
k^{\textrm{HTST}}=\nu \exp \left[-\frac{E_{\textrm SP}-E_{\textrm R}}{k_{\textrm b}T} \right]\\
\nu=\frac{\prod^f_i\nu_{{\textrm R},i}}{\prod^{f-1}_i\nu_{{\textrm SP},i}}
\end{eqnarray}
where ${E_{SP}}$ is the energy of the saddle point, ${E_R}$ is the energy of the minimum corresponding to the initial 
configuration,
$f$ is the number of degrees of freedom in the system, and 
${\nu_{SP,i}}$ and ${\nu_{R,i}}$ are 
frequencies of vibrational modes at the saddle point 
(excluding the unstable mode, the one corresponding to the negative eigenvalue) and the reactant configuration.
Several saddle point searches were carried out for each minimum visited 
until a confidence level of 0.99 was obtained as defined by Xu {\it et al.}~\cite{Xu08}.
Then, the simulation proceeded according to the traditional KMC algorithm by picking one of the
saddle points with probability proportional to the relative rates and advancing the simulated time by
\begin{eqnarray}
\label{equ:TimeStep}
\Delta t=-\frac{\ln \mu }{\sum_j k^{\textrm{HTST}}_j},
\end{eqnarray}
Here, ${\mu}$ is a random number in the interval $(0,1]$ and $j$ runs over all saddle points found on the energy ridge surrounding
this minimum. 
After picking a transition and advancing the clock, the state of the system corresponds to the minimum on the other side of the saddle point.  

The initial displacement for each saddle point search involved rotating and translating the H$_2$O admolecule while keeping the 
molecular geometry unchanged. The magnitude of the displacements was determined by values drawn from a Gaussian distributions with standard deviations of 0.25 radians and 0.25 \AA.  
On average a total of ca. 500 saddle point searches were carried out for each minimum visited.
Even though only the admolecule is displaced initially, some surface and subsurface molecules also move during the 
transitions.  Furthermore, annealing events that change the surface morphology are occasionally observed.
A study of the annealing events has been carried out and reported separately~\cite{Pedersen14}.
There, comparison was also made between results obtained from calculations using the TIP4P/2005f potential and DFT calculations.
Here, we focus on the admolecule diffusion.

While a large speedup is gained by skipping the vibrational motion of the atoms and  
focusing on the rare, activated events in the AKMC simulations,  
the presence of fast transitions, corresponding to low activation energy, 
will reduce the time increment in an iteration to a small value.  Even just one large 
rate constant for a possible transition in the system brings $\Delta t$ in eq 1 to a small value.
In a system where there is a wide range of energy barriers as in the case of disordered systems, here
the proton-disordered ice crystal, small energy barriers are likely to be present.
An essential component of the simulations conducted here was a systematic coarse-graining of the 
energy landscape where minima separated by low energy barriers 
were grouped together into a single state in the list of transitions, 
while maintaining the correct estimate of the residence time and 
escape transition mechanisms from this composite state~\cite{Pedersen12,Jonsson11}.  
The coarse-graining was limited here to include at maximum four minima to ensure sufficient resolution for determining
the mean squared displacement of the admolecule. The implementation of this algorithm in the EON software, however,
allows for an arbitrary number of minima to be included in a coarse-grained state~\cite{Chill14,Pedersen12}.

The AKMC simulations typically covered time intervals on the order of 0.1 $\mu$s 
as determined from the sum of time increments given by eq 3. 
The simulated diffusion paths extended throughout the surface, that is from one side
of the simulation box to the other, and thereby throughout the periodically replicated surface. 
From these paths, the possible binding sites of the admolecule were determined and rates of likely transitions involving 
admolecule hops between the sites as well as structural changes of the substrate.  
The AKMC simulation is, however, computationally demanding and the time intervals simulated here are 
only marginal for extracting statistically significant values of the diffusion coefficient.  
For the Fletcher surfaces (with and without defects) 
the information obtained in the AKMC simulations on binding sites and transition rates was stored and reused.
The data was resampled in
subsequent KMC simulations (without saddle point searches) covering longer time scale, from a few milliseconds at 200 K 
to tens of seconds at 100 K,
to obtain more accurate estimates of the mean squared displacement and, thereby, more accurate values of the diffusivity.  
Transitions involving changes in the surface configurations were not included in the KMC simulations (unlike the AKMC simulations of the proton-disordered surface).
When a transition involving substrate rearrangement was drawn from the table of events, the simulation was restarted
from a state where the surface had not reconstructed.
In this way, the diffusion contants for a given surface structure (such as a Fletcher surface with a specified 
number of defects) could be determined with good statistical sampling.
Such simulations were carried out at 100~K, 125~K, 150~K, 175~K and 200~K including $10^{8}$ KMC steps
using the adsorption sites, values of activation energy and pre-exponential factor obtained 
for each transition in AKMC simulations at 200 K.


\section{III. Results}
\label{sec:Results}

We first discuss the adsorption energy of the H$_2$O admolecule at sites identified from the AKMC simulations, then the 
activation energy for diffusion hops, and, finally, the rate of diffusion.
Results were obtained for the proton-disordered surface as well as Fletcher surfaces with various numbers of broken DH rows.

\subsection{Adsorption energy}
\label{sec:adsorption}

Due to the proton disorder of the ice Ih crystal, no two sites of the H$_2$O admolecule on the surface are the same, 
even for the Fletcher surface where the DHs form regular rows on the surface.  
The long-range electrostatic interaction from the proton-disordered crystal leads to 
widely ranging binding energy for the admolecule.
Fig.~2 shows the distribution for the proton-disordered surface and the DH-ordered Fletcher surface. 
In both cases, the distribution is broad and, surprisingly, is even broader for the ordered Fletcher surface
than the proton-disordered surface. 
On average, the binding energy of the admolecule is
larger on the proton-disordered surface, consistent with the fact that it has higher surface energy.

The range in adsorption energy is so broad that a significant number of sites provide higher binding energy than 
the cohesive energy of the ice Ih crystal, 0.64 eV (obtained for this interaction potential function). 
For the proton-disordered surface, these strong binding sites are 4\% of all the sites visited.
This result is consistent with the findings of BJ~\cite{Batista01,Batista15}. At such strong binding sites,
the admolecule can form three strained hydrogen bonds with the substrate, while the cohesive energy corresponds to two
hydrogen bonds~\cite{Batista01}.
Admolecules sitting in these sites will be highly stable and have little tendency to migrate to kink sites 
where the binding energy will, on average, necessarily be equal to the cohesive energy.  
There will, of course, also be a variety of binding sites at kinks, with a broad distribution of the 
values of the binding energy, but the most probable
state of a surface with an incomplete surface layer will have H$_2$O admolecules sitting at strongly binding sites on the 
terrace as well as at strongly binding kink sites.  
The time-averaged binding energy of the admolecule from the simulated path at 200 K was, indeed, found to be 
larger than the cohesive energy, 0.67 eV, on the proton-disordered surface showing that the admolecule spends most of its 
time in the strongly binding sites.  On the Fletcher surface, the time-averaged binding energy at the same temperature 
was also large, 0.61 eV.

\begin{figure}
\centering
\includegraphics[width=.65\textwidth]{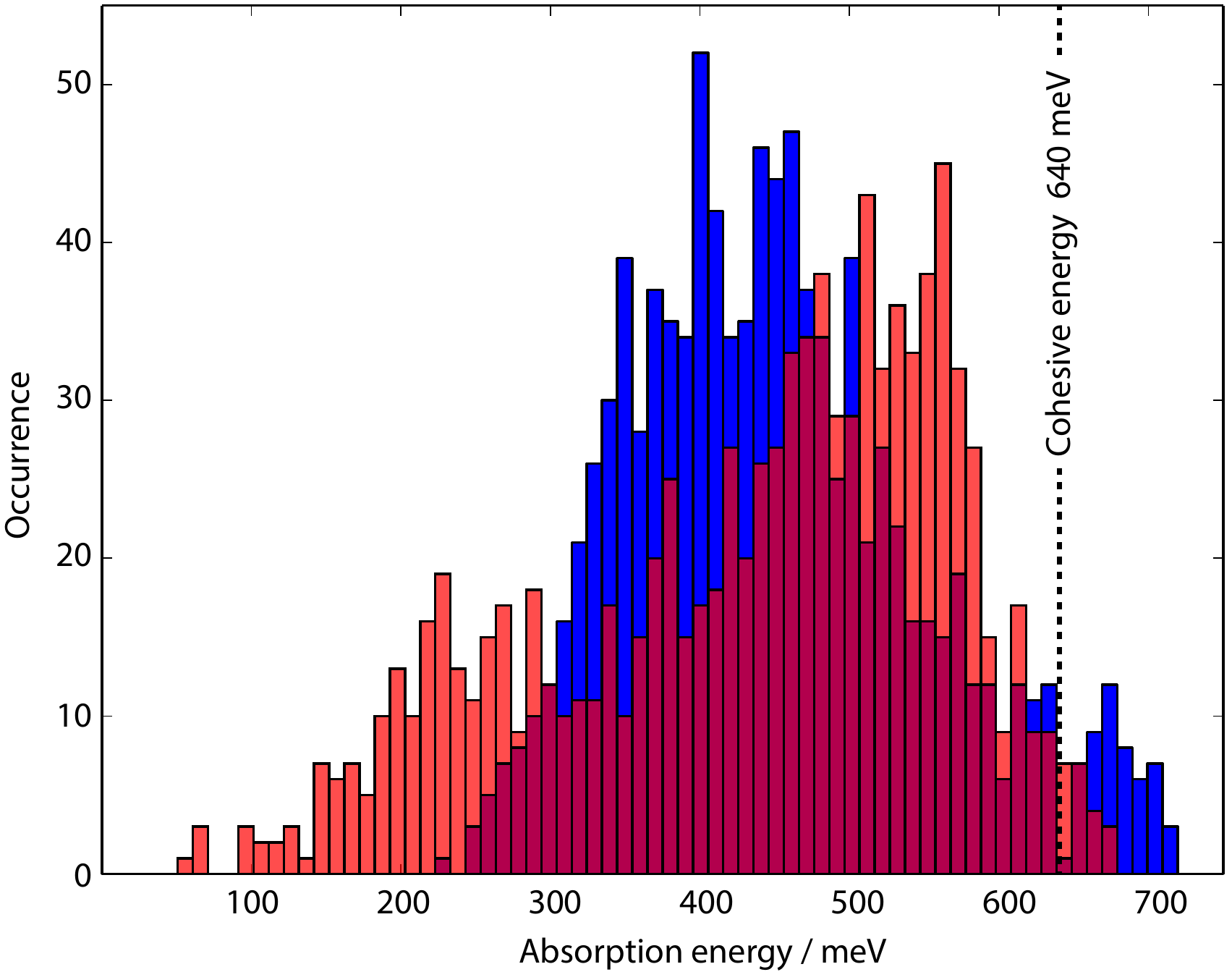}
\caption{
Distribution of adsorption energy values for a H$_2$O admolecule on a Fletcher surface (light red) and a 
surface with disordered arrangement of dangling H atoms (blue). (Dark red represents overlap of the two histograms).
The sites were visited during the long-timescale adaptive kinetic Monte Carlo simulation.
About 4\% of the sites on the disordered surface have 
higher binding energy than the cohesive energy of ice Ih, which was calculated to be 0.64 eV using the 
TIP4P/2005f interaction potential function. 
On average the binding energy is larger on the disordered surface, consistent with higher surface energy.
}
\label{fig:binding}
\end{figure}

\subsection{Diffusion hops}
\label{sec:activation}

The activation energy for diffusion hops of the H$_2$O admolecule on the proton-disordered surface also 
turns out to have a wide range.
An AKMC simulation was carried out for diffusion of the admolecule at 175~K.
A time interval of 313 ms was simulated with a total of 90.000 AKMC iterations. 
Fig.~3 shows the distibution of activation energy obtained from the simulated path. 
Most of the transitions have a low activation energy, below 0.1~eV, 
but a significant number of transitions have an activation energy between 0.1 and 0.2~eV.  
Transitions with even higher activation energy were occasionally also selected to be part of the AKMC path, but only rarely. 
There is a finite probability that the random number drawn to select the next transition in the path points to a mechanism that involves
relatively high activation energy.  The coarse-graining where minima separated by a low energy barrier are grouped 
together in a single state also reduces the occurrence of transitions with low activation energy.
When two or more minima have been grouped together in the coarse-graining, the activation energy of an escape event
is recorded as the energy of the saddle point minus the energy of the lowest minimum in the group.

The mechanism of the diffusion hops is similar to what has been presented previously by BJ. 
As the admolecule moves from one site to another,  the underlying surface molecules typically rotate to maintain hydrogen bonding
as much as possible. It is, therefore, extremely 
important not to artificially constrain the surface molecules and allow several layers in the ice lattice to relax during the 
climb up the potential surface to saddle points~\cite{Batista01}. 
The diffusion hops did not, however, involve concerted displacements where
the admolecule replaced a surface molecule, a mechanism that has been found to be important in 
metal adatom diffusion~\cite{Jonsson11,Feibelman90,Villarba94}.
Such concerted displacement events have, nevertheless, been found to be the prevailing mechanism of annealing events in
the surface layer of ice Ih~\cite{Pedersen14}.

The pre-exponential factor in the HTST expression, eq 2,  for the rate constant of diffusion hops was evaluated
from the vibrational frequencies obtained by constructing the Hessian matrix for all degrees of freedom in the system.
Calculations based on a sub-matrix, including only the atoms displaced most, turned out not to satisfy detailed balance well 
enough to be applied in the KMC resampling simulations. 
A slight drift in the diffusion paths resulted from a violation of detailed balance.
The prefactors were typically found to be on the order of $10^{13}$ s$^{-1}$.

\begin{figure}
\centering
\includegraphics[width=.65\textwidth]{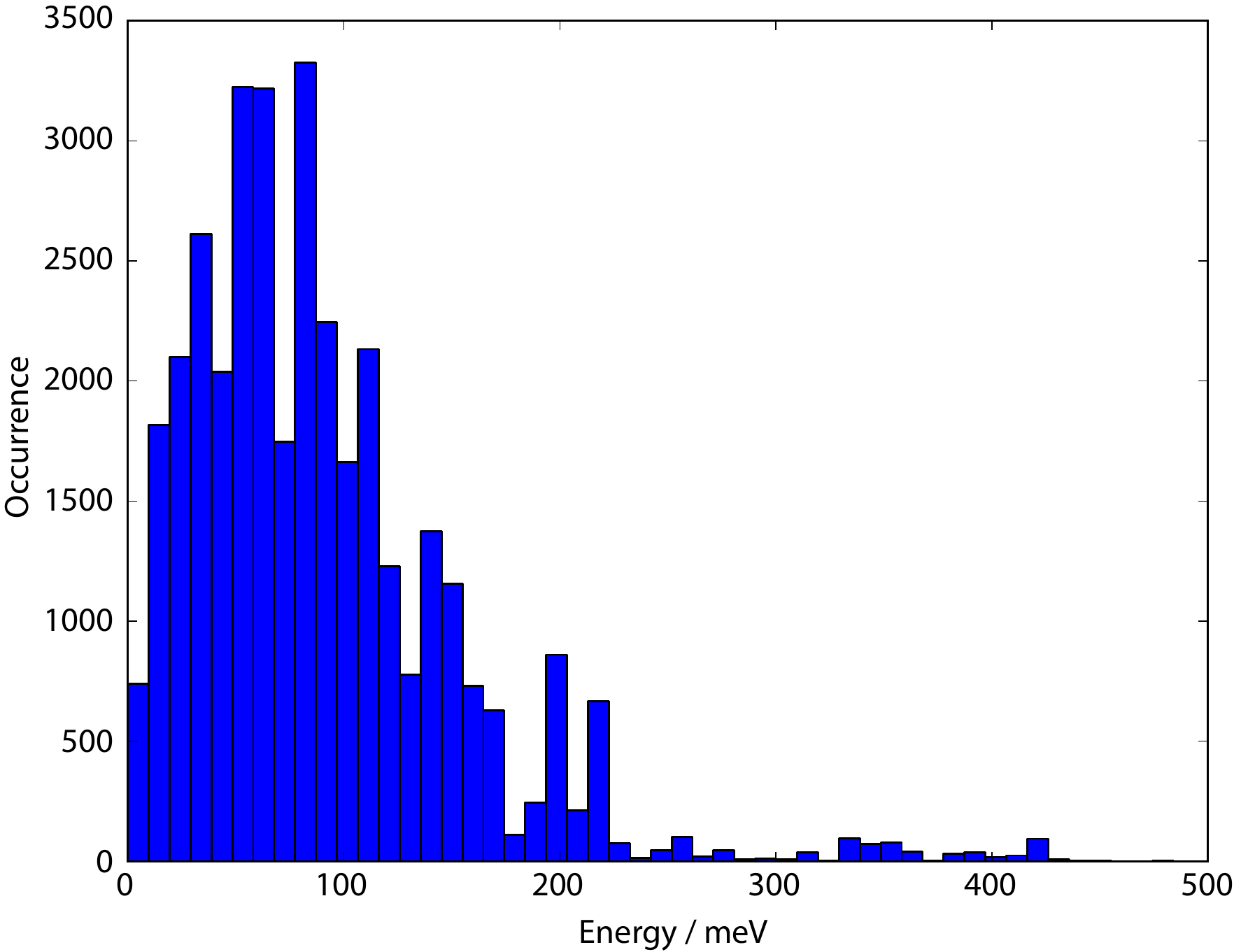}
\caption{Distribution of activation energy values for H$_2$O admolecule diffusion hops on the proton-disordered surface at 175 K.
The values where obtained from long-timescale adaptive kinetic Monte Carlo simulations of the diffusion spanning 313 ms. 
While most of the values are below 0.1 eV, transitions with activation energy over 0.2 eV turn out to be essential for 
obtaining a diffusion path extending over the full width of the simulated surface.
}

\label{fig:Barriers}
\end{figure}


\subsection{Diffusion constants}
\label{sec:diffusion}

The mean squared displacement was calculated from the 
AKMC simulated diffusion path on the proton-disordered surface described above.
The squared displacement was calculated over an interval of 100 ns for several different choices of the origin.
The mean squared displacement obtained in this way is shown in Fig.~4.
While the statistical fluctuations are large, the long time results can be fitted to a straight line to estimate a diffusion 
coefficient according to the Einstein-Smoluchowski equation 
\begin{eqnarray}
\label{equ:Diffusivity}
D&= \frac{\langle{ \vert{ {\bf r}(t_0+\tau)-{\bf r}(t_0)}\vert^{2}}\rangle} {2d \tau},
\end{eqnarray}
where $d$ is the number of dimensions, being $d=2$ in the present case.
This gives a diffusion coefficient of $D ~=~ (6 \pm 2) \cdot 10^{-10} $~cm$^2$/s, the estimate of the error bar being obtained 
from an upper and lower bound on the slope of a line fitting the data.
%

\begin{figure}
\centering
\includegraphics[width=.65\textwidth]{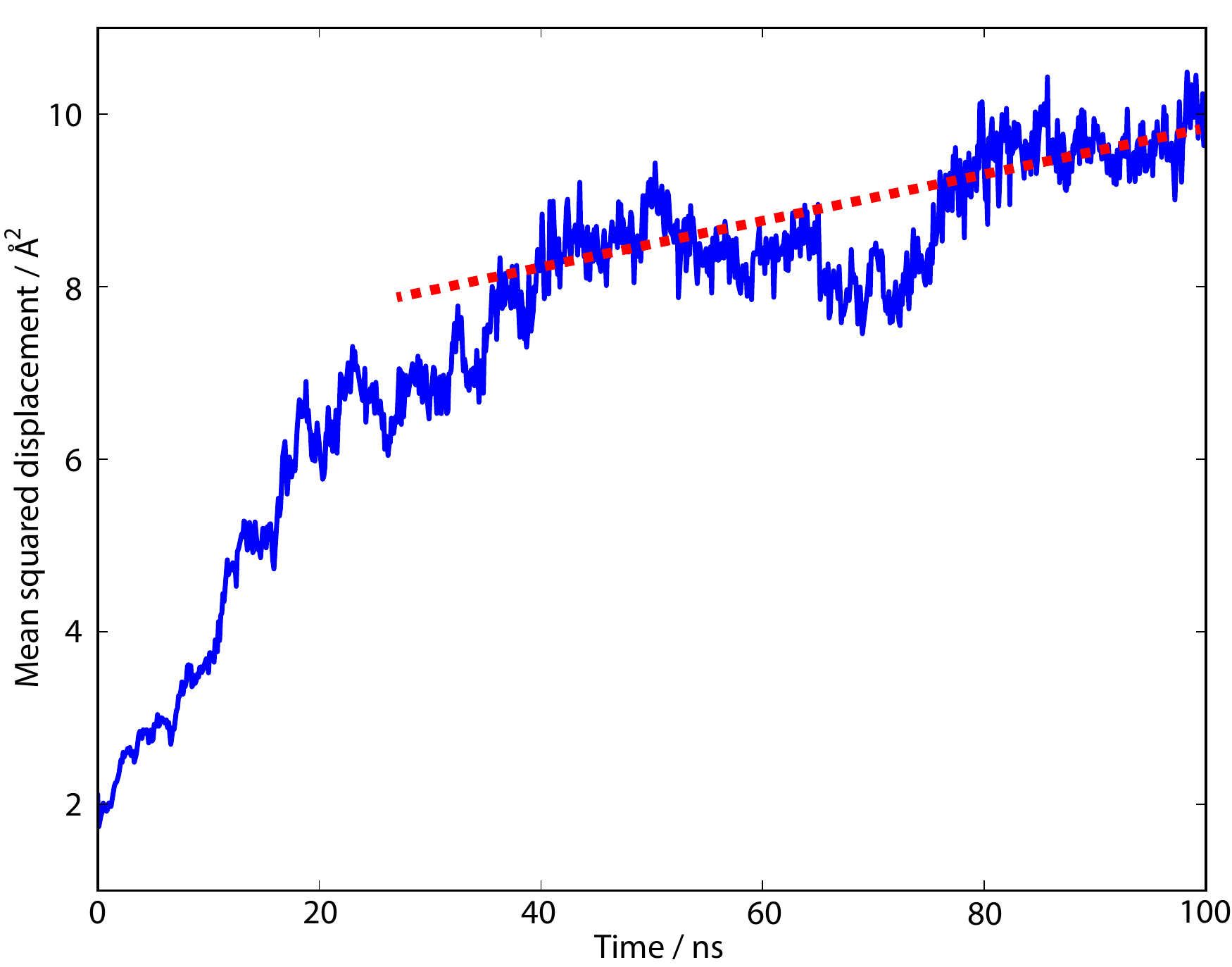}
\caption{
Mean squared displacement of an H$_2$O admolecule on the proton-disordered surface as a function of time
obtained from the adaptive kinetic Monte Carlo simulation spanning 313 ms time interval at 175 K.  From the long time slope
of the curve in the range from 40 to 100 ns, a diffusion coefficient of $D$~=~(6 $\pm$ 2)$\cdot$10$^{-10}$ cm$^2$/s was estimated.
}
\label{fig:Diffusivity}
\end{figure}

An inspection of the surface before and after the diffusion simulation reveals that significant changes have occurred 
in the DH pattern, as shown in Fig.~5.  While the ordering of the original disordered surface has a cluster of DHs
with one of them surrounded by five DHs at near neighbor sites, 
the final configuration has a more linear ordering where there are at most three near neighbor DHs. 
This can be seen as an evolution of the proton-disordered surface towards a Fletcher-type surface.
Most of these annealing transitions occurred early on in the simulation and do not affect the long range linear fit 
used to determine the diffusion constant.
While the initial states in the MMF saddle point searches only involve 
displacement of the admolecule, the surface molecules (as well as molecules in layers further down from the surface) are 
free to move and rotate and in some of the transitions the DH pattern on the surface changed.  
Such annealing transitions have been discussed previously~\cite{Pedersen14}. 
The focus here is on the H$_2$O admolecule binding and diffusion.  

\begin{figure}
\centering
\includegraphics[width=.35\textwidth]{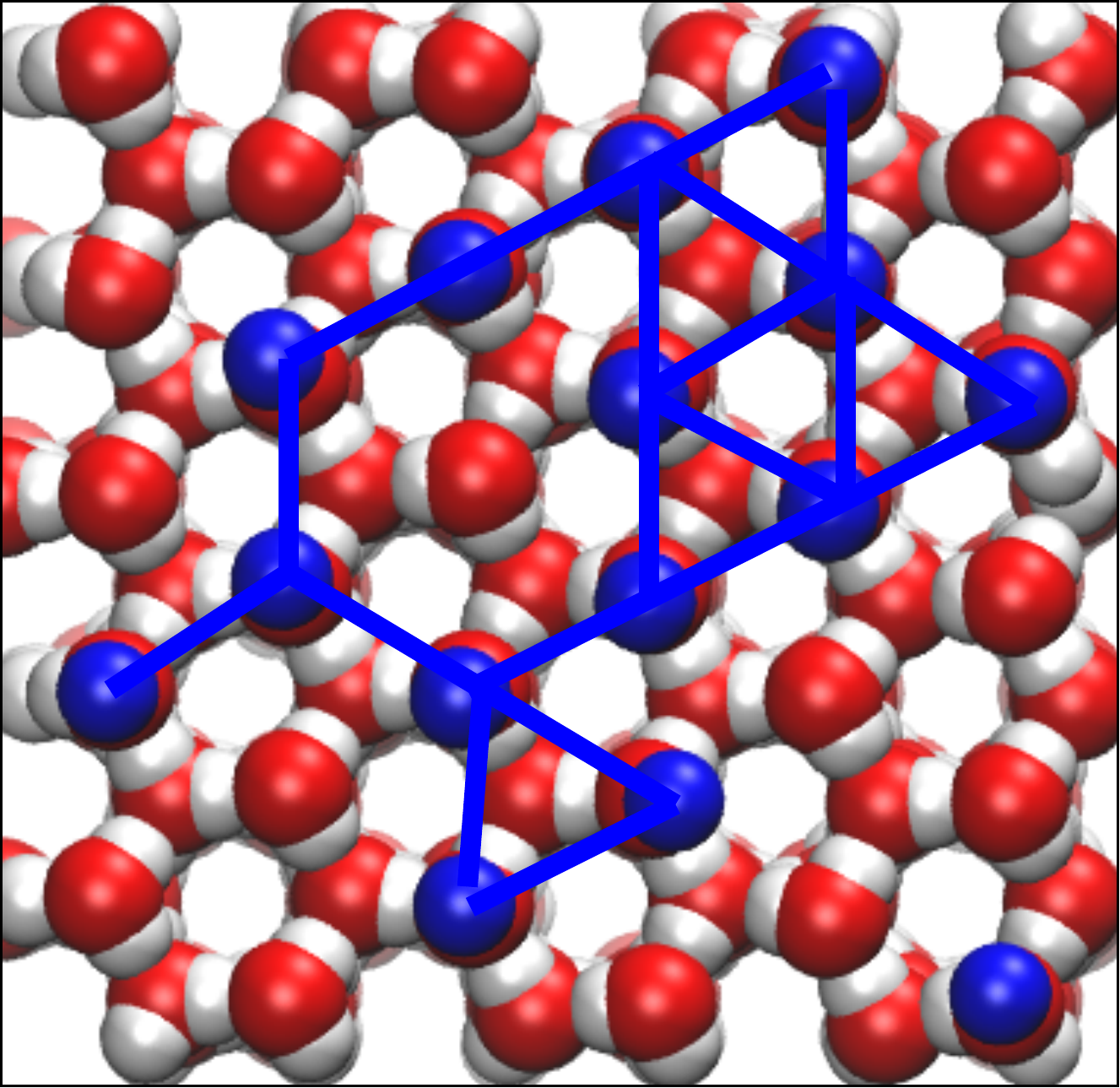}
\includegraphics[width=.35\textwidth]{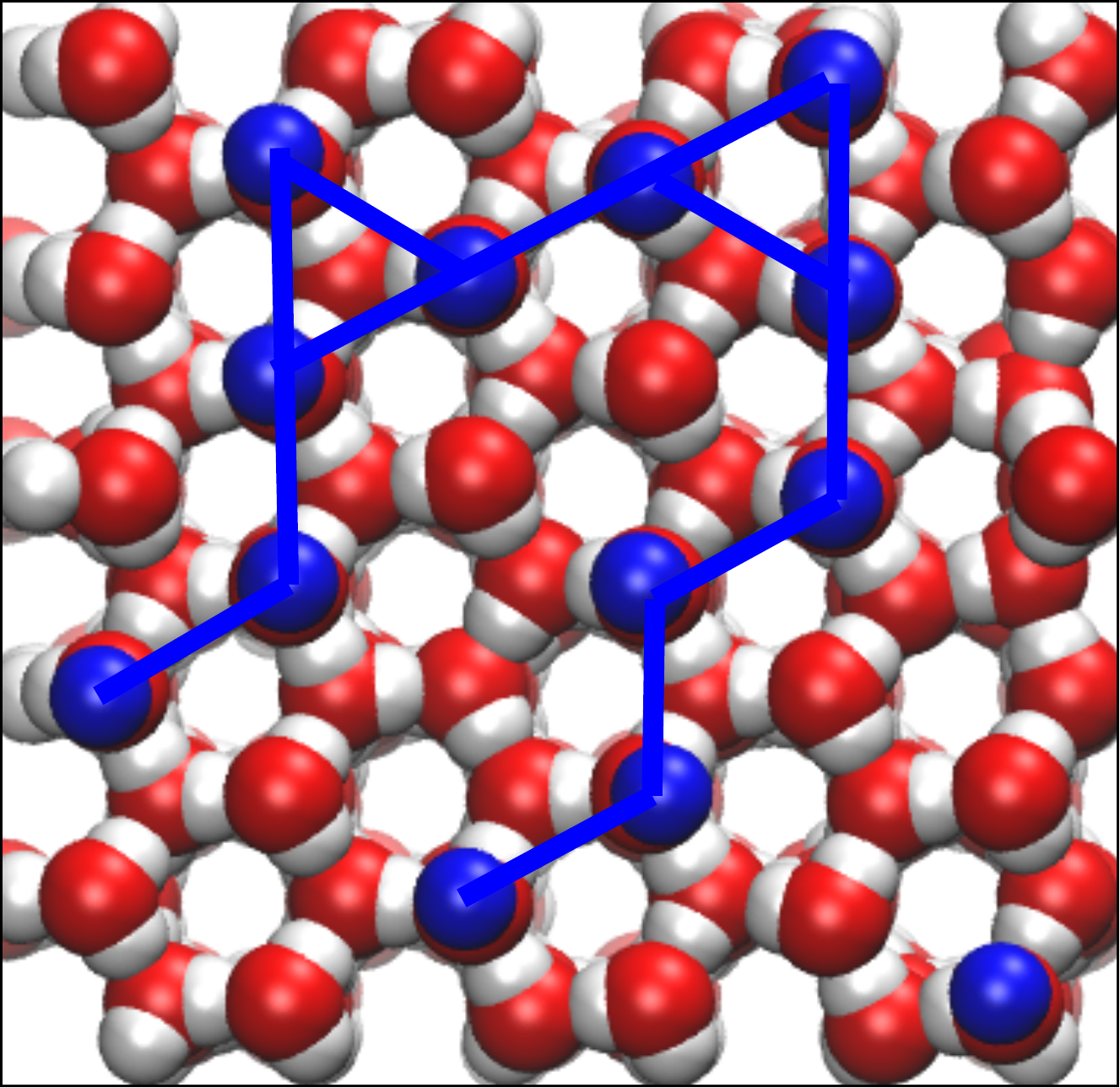}
\caption{
Arrangement of dangling H atoms before (left) and after (right) an adaptive kinetic Monte Carlo simulation of 
H$_2$O admolecule diffusion on the proton-disordered surface.  
The color code is the same as in Fig.~1.
A blue line is drawn between dangling H atoms located on adjacent sites.  
The initial clustering of the dangling H atoms is reduced during the diffusion simulation as the 
surface energy is lowered by spacing them further apart.
Since the substrate molecules are free to move in the saddle point searches, surface rearrangement processes 
are included in the table of possible transitions, even though only the admolecule is initially displaced.
}
\label{fig:Dbondreararrangement}
\end{figure}

Similar simulations were carried out for the Fletcher surface and defected Fletcher surfaces (shown in Fig.~1).
Also there, annealing events that change the DH ordering occasionally occur. In order to extract accurate 
diffusion constants for surfaces with a certain, well defined arrangement of DHs, the annealing events were excluded from the 
table of possible transitions. Also, in order to obtain better statistics for the mean squared displacement and the 
deduced diffusion constant, regular KMC simulations were carried out for long time intervals after mapping out the
possible adsorption sites and likely diffusion hops in AKMC simulations. 
Two such paths simulated at 150~K are shown in Fig.~6 where the admolecule
has traveled over distances on the scale of micrometers. The anisotropy of the diffusion on the Fletcher surface is evident from 
the figure, the diffusion being faster and the paths extending further in the direction along DH rows. 
In order to characterize the anisotropy of the diffusion on the Fletcher surface, the mean squared displacement parallel and 
perpendicular to the DH rows were calculated separately and the long time results fitted by a straight line to 
extract a diffusion constant using eq 3 with $d=1$. 
Fig.~7 shows an example of such a calculation, the 
mean squared displacement along rows obtained from three independent simulations at 100~K. 
For each curve a linear fit of the long time values was used to extract a diffusion constant and then an average taken of the 
values obtained from the three curves.  By calculating separately the mean squared displacement parallel and perpendicular 
to the rows, a parallel,  $D_\parallel$, and a perpendicular, $D_\perp$, diffusion constants were obtained. 

\begin{figure}[t]
\centering
\includegraphics[width=.7\textwidth]{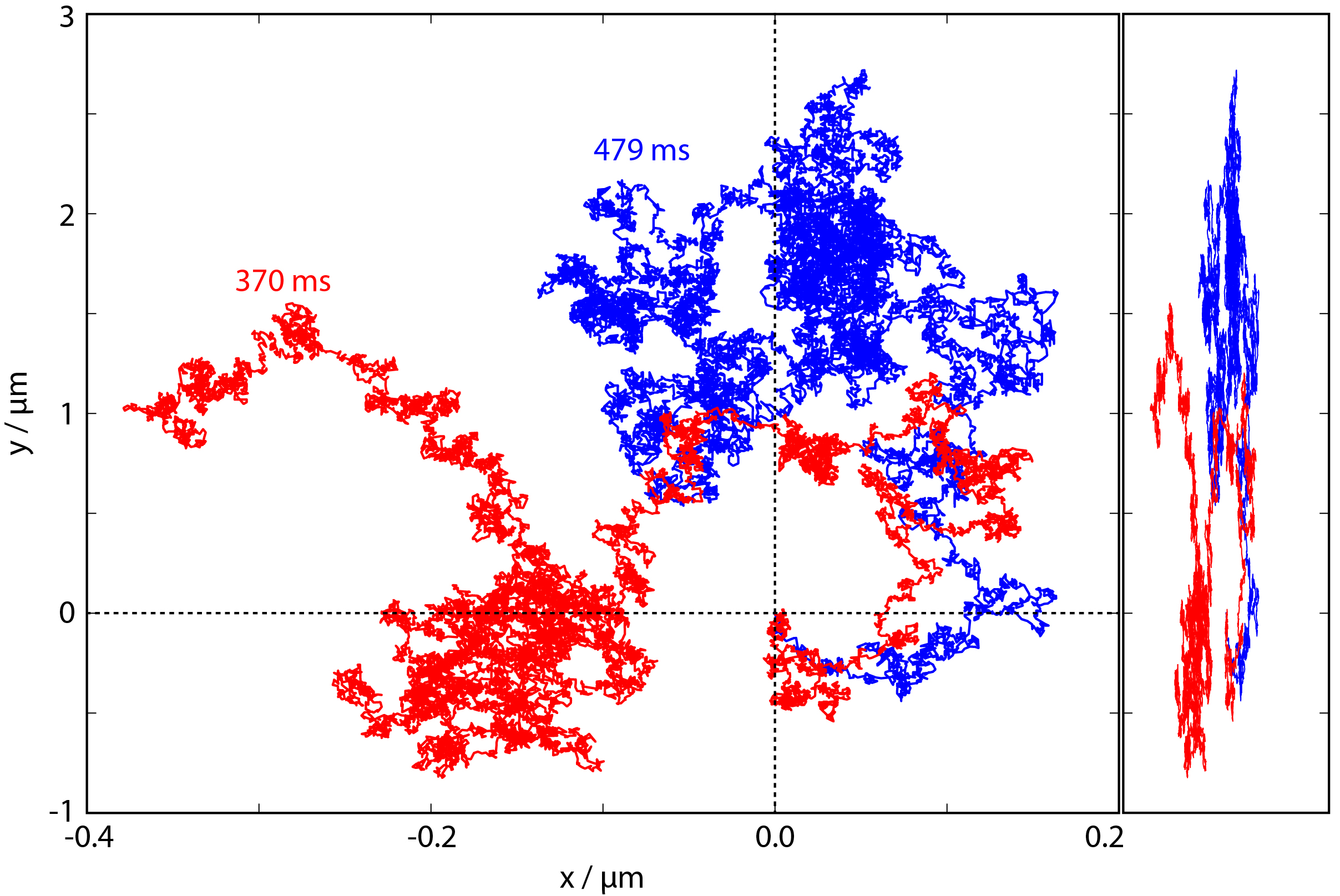}
\caption{
Two simulated diffusion paths of an H$_2$O admolecule on a Fletcher surface with intact rows of dangling H atoms at 150~K.
In the left figure, the scale of the graph corresponding to the distance perpendicular to the rows, $x$, is chosen to 
be different from the one parallel to the rows, $y$, to clearly display the trajectories.
In the right figure, the scale is the same and the strong anisotropy of the diffusion can be seen clearly,
The diffusion rate along the rows is 70 times higher than perpendicular to the rows at this temperature.
}
\label{fig:DiffPath}
\end{figure}

Fig.~8 shows results obtained for the diffusion constant $D_\parallel$ at 
various temperature values ranging from 100 to 200~K.
The diffusion constant is found to vary with temperature according to the Arrhenius equation and the slope of the 
lines gives the effective activation energy of diffusion in each of the two directions.
The activation energy for diffusion is found to be 0.22~eV for diffusion perpendicular to the rows,
but 0.16~eV for diffusion parallel to the rows.
The values are listed in Table 1,
along with the parallel diffusion constant and ratio of the diffusion constants, $r=D_\parallel/D_\perp$, at 100 and 200~K.
These results show that the diffusion on the Fletcher surface is highly anisotropic,
being three orders of magnitude faster along the DH rows than perpendicular to the rows
at 100~K and an order of magnitude faster at 200~K. 

\begin{figure}
\centering
\includegraphics[width=.65\textwidth]{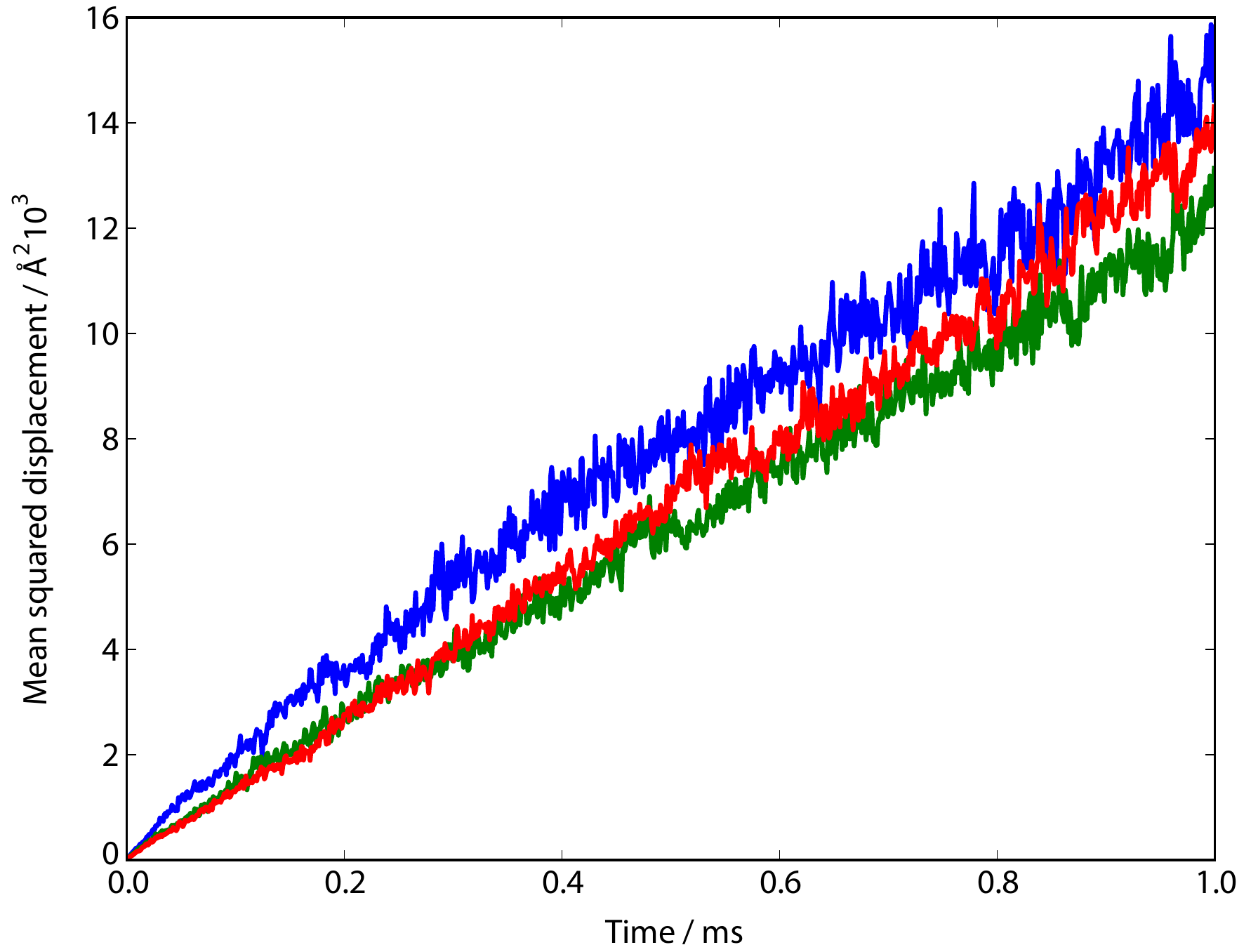}
\caption{Mean squared displacement along rows of dangling H atoms of the Fletcher surface obtained from 
simulations corresponding to a temperature of 100~K.  Results of three independent calculations are shown. 
The average diffusion constant deduced from these simulations is shown in Fig.~8 
along with analogous results for higher temperature and for defected Fletcher surfaces.
}
\label{fig:Diffat100K}
\end{figure}

\begin{figure}
\centering
\includegraphics[width=.65\textwidth]{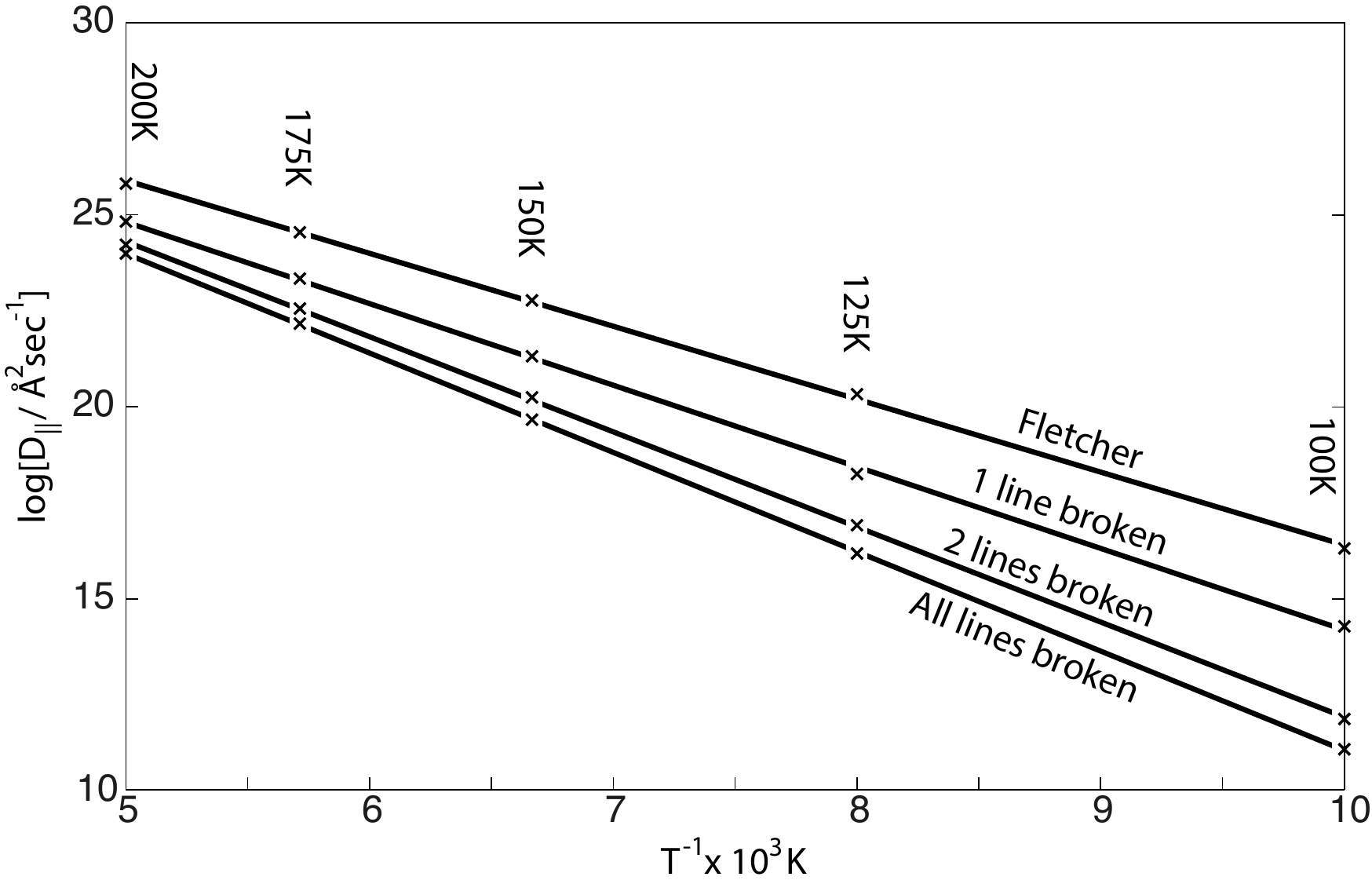}
\caption{
Diffusion coefficient obtained from the mean squared displacement of the admolecule parallel to the rows of dangling
H atoms on the Fletcher surface with regular, intact rows as well as defected Fletcher
surfaces where one or more dangling H atom is placed in between the rows (as shown in Fig.~1)
as a function of inverse temperature over the range 100 to 200~K.  
}
\label{fig:Arrhenius}
\end{figure}


\begin{table}
\begin{center}

\caption{
Activation energy of surface diffusion of an H$_2$O molecule on ice Ih(0001) surface, 
parallel diffusion constant $D_\parallel$ at 100~K, 
and ratio of diffusion rate parallel and perpendicular, $r=D_\parallel/D_\perp$ to rows of 
dangling H atoms on the perfect and distorted Fletcher surface models.
The activation energy was obtained from the temperature dependence of the diffusivity over a temperature range from 100 to 200~K.
The ratio, $r$, is given for the highest and lowest temperature.
}
\vskip 0.3 true cm
\begin{tabular}{l | c c c c c c }
  		         & E$^a_\perp$/eV & E$^a_\parallel$/eV & $D_{\parallel,100\rm K} \cdot 10^{10}$/(cm$^2$/s)  &  r$_{100 \rm K}$ &  r$_{200 \rm K}$ \\
\hline
Fletcher 		&  0.22  & 0.16  &    6.1      &    570 & 24 \\
1 row broken	&  0.23  & 0.18  &    0.80    &    180 & 13 \\
2 rows broken 	&  0.25  & 0.21  &    0.072   &    16   & 4  \\
all rows broken 	&  0.22  & 0.22  &    0.047   &      1   & 1
\label{tab:barriers}
\end{tabular}
\end{center}
\end{table}


The perfect Fletcher surface is, however, not the most stable arrangement of the DHs.
A slighlty lower energy ordering was found by breaking one of the rows and placing a DH in between rows, as shown in Fig.~1.
Such defective Fletcher surfaces have, furthermore, larger entropy and will, thereby, have lower free energy at finite temperature
than the perfect Fletcher surface. We now address how such defects affect the diffusivity. 
Results of simulations of diffusion on Fletcher surfaces with one, two and three broken rows ({\it i.~e.}, all three rows in the
simulation cell) are shown in Fig.~8.  Again, clear Arrhenius dependence on temperature is obtained and 
the extracted values of the activation energy are listed in Table 1. The activation energy for diffusion along rows 
on the perfect Fletcher surface increases as more of the rows are broken and DHs placed between the rows.
When all rows have been broken, the activation energy for diffusion parallel to the rows equals the activation energy for
diffusion perpendicular to the rows. The anisotropy is, thereby, eliminated by just a single DH between each 
pair of adjacent rows on this model surface.

%
\section{IV. Discussion and summary}
\label{sec:Discussion}

The study presented here of H$_2$O admolecule diffusion on the ice Ih(0001) surface improves upon and extends the
early simulation study of BJ~\cite{Batista01}.  
A more sophisticated simulation methodology is used, 
regarding both the simulation algorithm and the interaction potential.
Also, the effect of DH ordering is studied, ranging from perfect 
rows of DHs on the Fletcher surface, to a Fletcher surface where all rows are broken, and a (annealed) proton-disordered surface. 
The results are consistent with published measurements of Brown and George in that the diffusion 
constant for the most likely models of the ice surface, an annealed proton-disordered surface or 
disordered Fletcher surface (all rows broken), is 
less than or similar to the upper bound set by the experimental results~\cite{Brown96}.

The adsorption energy is found to vary greatly from one site to another due to the proton disorder in the ice lattice, 
with ca. $4 \%$ of the sites on the proton-disordered surface providing larger binding energy than the cohesive energy
of the crystal. 
Even the perfectly ordered Fletcher surface has a broad distribution in adsorption energy because of the long range electrostatic 
interaction as well as the variation in the number of DHs at the nearest molecules in the surface layer. 
Admolecules sitting in these strongly binding sites
will be highly stable and not have a tendency to migrate to kink sites, 
where the binding energy will, on average, necessarily be equal to the cohesive energy.   
The binding energy of an H$_2$O molecule at kink sites will, of course, also vary greatly, but the most common
situation of a surface with an incomplete surface layer will have admolecules sitting at strongly binding kink sites as well as strongly binding sites on the flat terrace.  
As a result, the terraces will not be free of admolecules, even when the temperature is high enough for diffusion from 
terrace sites to step and kink sites to be possible. 
It has been suggested~\cite{Batista01}
that this could explain the observation of dispersionless modes in inelastic He-atom scattering~\cite{Glebov00} 
since the vibration of isolated admolecules is likely to couple only weakly to the vibrations of the crystal lattice. 

The diffusion is found to be highly anisotropic on the perfect Fletcher surface, with the ratio of diffusion parallel and 
perpendicular to the rows ranging from 24 at 200~K to 570 at 100~K.  
However, only a slight disordering of the Fletcher surface, obtained by placing a few DHs in between the rows, 
a likely configuration at finite temperature because of both energetic and entropic considerations,
eliminates the anisotropy.
These findings indicate that it will be difficult to observe experimentally an anisotropy in admolecule diffusion on an Ih(0001) 
surface even if the DHs are close to being ordered in Fletcher rows. Only a slight deviation from the perfect order leads 
to isotropic diffusion.

The effective activation energy for diffusion on the Fletcher surface with all rows broken was
determined to be 0.22~eV from the variation of the diffusion constant over a temperature range of 100 to 200~K. 
The time averaged binding energy of the H$_2$O molecule on the surface calculated over a diffusion path
obtained at 200~K was found to be 0.61~eV.  The ratio of the diffusion activation energy to the desorption activation energy is, 
therefore, found to be nearly 1/3.  This is an important quantity that enters, for example, in the modeling of ice crystal growth.

The quantitative accuracy of the results obtained here could be improved by using a more accurate description of the molecular
interaction.
The simulations presented here made use of a simple, point charge model
fitted to bulk properties, so its applicability to surface properties can be questioned.
Further work, using a more accurate description, such as the 
single center multipole expansion (SCME) potential~\cite{Wikfeldt13} would be desirable.  
The SCME potential gives better agreement with the experimental lattice constants and cohesive energy of ice Ih 
than the potential function used here.
The multipole expansion used in the SCME potential also gives electrostatics 
in good agreement with {\it ab initio} and density functional theory
calculations while point charge models do not~\cite{Batista99b}. 
It would, therefore, be interesting to repeat the types of simulations carried out here with the SCME potential.

It would also be interesting to assess whether the inclusion of quantum mechanical effects would alter the results obtained here.  
The simulations presented here were based on a purely classical description of the atoms.  
The binding energy and the transition rates can be affected by quantum delocalization of the atoms, especially the H atoms.  
Such a simulation could be carried out using harmonic quantum transition state theory where zero point motion
and quantum tunneling is taken into account. 
The AKMC method can, in principle, be extended in this way and simulations can be carried out 
where the effect of quantum tunneling on the transition rates is included~\cite{Jonsson11,Schenter94}.


\begin{acknowledgement}
A.P. and H.J. were supported by the Icelandic Research Fund, the University of Iceland Research Fund and
the Academy of Finland (grant 263294). 
L.J.K and H.M.C were suppored by the European Research Council (ERC-2010-StG, Grant Agreement No. 259510-KISMOL). H.M.C. thanks The Netherlands Organization for Scientific Research (NWO) (VIDI 700.10.427),
A.P. and H.J. thank Jean-Claude Berthet for helpful discussions, and 
A.P. thanks COST Action CM0805 for travel funds. 
\end{acknowledgement}


\end{document}